# QUANTUM DIFFUSION AS A PROCESS OF LINEAR QUANTUM DYNAMICS

K.E. Eriksson, Complex Systems Group, Department of Space, Earth and Environment, Chalmers University of Technology, SE 41296, Gothenburg, Sweden. frtkee AT chalmers.se

Abstract

Quantum diffusion, as developed in the 1990s, could explain how a system, subject to measurement, goes into an eigenstate of the measured observable. Here it is shown that quantum diffusion theory can be interpreted as a result within linear relativistic quantum mechanics. Thus, in contrast to what is widely believed, quantum measurement can be analyzed within the theory of quantum mechanics itself.

## 1. Introduction

An early paper to suggest a mechanism, how a system, subject to measurement, goes into an eigenstate of the measured observable, is a 1992 paper on quantum diffusion by N. Gisin and I.C. Percival [1]. Their basic equation of motion is a highly non-linear generalized Schrödinger equation (with stochastic variables) for the state of the measured system.

We show here that the equation of motion of Gisin and Percival, transformed into an equation for the density matrix of the measured system, can be interpreted as an equation of ordinary linear quantum mechanics for the measured system and part of the measurement apparatus. Thus, the paper by Gisin and Percival can be understood as an early treatment of quantum measurement within quantum mechanics itself. Quantum mechanics is a rich theory, and quantum measurement can be viewed and analyzed as a physical process within quantum theory.

In this sense, the theory of quantum diffusion can be viewed as an answer to Albert Einstein's criticism of the view of quantum mechanics as an inherently probabilistic theory. The source of randomness is instead unknown details in the part of the measuring apparatus first met by the measured system.

The potential relevance of the work of Gisin and Percival for understanding measurement, was observed by J. Bell [2] and by A. Whitaker [3], but these reviewers considered the work as a generalization of ordinary quantum mechanics rather than as a consequence of the linear theory.

A quantum-mechanical reasoning led Gisin and Percival to their basic non-linear Schrödinger equation. We have not evaluated to what extent this reasoning can be considered as a formal proof. Our work can however be viewed as an independent support of the soundness of their arguments. Here the Gisin-Percival paper [1] is treated as a representative work of the development in quantum diffusion and other papers in the field around the same time [4] have not been discussed.

In earlier work [5,6], we have argued for quantum mechanics as a complete theory, rich enough for an analysis of measurement. These works have been relatively close to quantum diffusion, but we have not analyzed the relationship between the two theories in detail before.

## 2. The equation of motion for quantum diffusion

We start with the stochastic equation of motion for quantum diffusion of a system $\mu$ as given by Gisin and Percival (Eq. (1.2) of [1]), a non-linear stochastic equation. However, here we omit the internal interaction Hamiltonian for the described system. We then have the equation:

$$|d\psi\rangle = \sum_m \left(2\langle L_m^\dagger \rangle_\psi L_m - L_m^\dagger L_m - \langle L_m^\dagger \rangle_\psi \langle L_m \rangle_\psi \right)|\psi\rangle dt +$$
$$+ \sum_m \left(L_m - \langle L_m \rangle_\psi\right)|\psi\rangle d\xi_m, \quad \langle L_m \rangle_\psi = \langle \psi | L_m | \psi \rangle.$$
(2.1)

The operators $L_m$ characterize the interaction with the medium met by the quantum system $\mu$ described by the state $|\psi\rangle$. The stochastic parameters $\xi_m$ have the statistical means,

$$\langle\langle \xi_m \rangle\rangle = 0, \quad \langle\langle \xi_m \xi_n \rangle\rangle = 0, \quad \langle\langle \xi_m \xi_n^* \rangle\rangle = 2\delta_{mn} dt.$$
(2.2)

The density matrix is the projection operator

$$\rho = |\psi\rangle\langle\psi|,$$
(2.3)

the differential of which is

$$d\rho = d|\psi\rangle\langle\psi| + |\psi\rangle d\langle\psi| + d|\psi\rangle d\langle\psi|.$$
(2.4)

In calculating this from (2.1), some second-order (i.e., $dt$) terms in $d|\psi\rangle\langle\psi| + |\psi\rangle d\langle\psi|$, cancel against terms in $d|\psi\rangle d\langle\psi|$, and

the resulting relation is of second degree in $\rho$:

$$d\rho = \sum_m \left(2L_m\rho L_m^\dagger - L_m^\dagger L_m \rho - \rho L_m^\dagger L_m\right)dt + \\ + \sum_m \left((L_m - \mathrm{Tr}(L_m\rho))\rho d\xi_m + \rho\left(L_m^\dagger - \mathrm{Tr}(\rho L_m^\dagger)\right)d\xi_m{}^*\right). \tag{2.5}$$

We shall use this as the equation of motion characterizing quantum diffusion.

We let the system $\mu$ be a particle which is scattered by an extended target $A$. We shall describe this <u>within linear quantum mechanics</u> [5,6]. We shall see that Equation (2.5) then follows from the linear theory as an equation of motion for $\mu$ during its passage of $A$.

## 3. <u>Scattering of a particle on an extended target</u>

In scattering theory, it is convenient to define from the unitary scattering operator $S$, a transition operator $M$, as follows

$$\langle f|(S-1)|i\rangle = \delta(P_f - P_i)\langle f|M|i\rangle. \tag{3.1}$$

where $P_i$ and $P_f$ are the energy-momentum vectors of the initial state $|i\rangle$ and the final state $|f\rangle$ of the scattering process. The transition rate from $|i\rangle$ to $|f\rangle$, suitably normalized, is

$$|\langle f|M|i\rangle|^2 = \mathrm{Tr}\left(|f\rangle\langle f|M\rho^{(i)}M^\dagger\right), \quad \rho^{(i)} = |i\rangle\langle i|, \tag{3.2}$$

Thus the (non-normalized) density operator for the final state, assumed to be orthogonal to $|i\rangle$ is

$$R = M\rho^{(i)}M^\dagger. \tag{3.3}$$

The total transition rate is

$$w = \sum_f |\langle f|M|i\rangle|^2 = \mathrm{Tr}R, \tag{3.4}$$

and the normalized final state is

$$\rho = \frac{1}{w}R = \frac{M\rho^{(i)}M^\dagger}{\text{Tr}(M\rho^{(i)}M^\dagger)}. \tag{3.5}$$

The crucial quantity is the transition operator $M$. We shall leave the strict plane-wave description of $\mu$ and $A$. Instead, we consider a wave-packet description with broad wave-packets. We assume the $\mu$ wave-packet to enter $A$ at the time $0$. We describe the process in the center-of-mass frame, practically coinciding with the rest frame of the larger system $A$. We define $M(t)$ to be the transition operator for $\mu A$-scattering with an interaction going on during the time interval $(0,t)$. We consider $\mu \cup A$ to be a closed system during this time. The larger system $A$ is unknown in its details and we shall study how it affects $\mu$ during the time period of interaction $(0,t)$. We assume the initial state of $\mu \cup A$ to be

$$|\Phi(0)\rangle = |\psi(0)\rangle_\mu \otimes |\alpha\rangle_A = \sum_m \psi_m(0)|m\rangle_\mu \otimes |\alpha\rangle_A. \tag{3.6}$$

The discussion will first be done for a given initial state $|\alpha\rangle_A$ for $A$. We write the state of $\mu \cup A$ at the final time $t$ of the interaction as

$$|\Phi(t)\rangle = M(t)|\Phi(0)\rangle = \sum_m \psi_m(t)|m\rangle_t;$$
$$|m\rangle_t = |m\rangle_\mu \otimes |\beta(\alpha,m,t)\rangle_A, \tag{3.7}$$

using an orthonormal basis for $\mu$,

$$_\mu\langle m|n\rangle_\mu = \delta_{mn}, \tag{3.8}$$

and define $|\beta(\alpha,m,t)\rangle_A$ to be the normalized state vector for $A$, entangled with $\mu$ in the state $|m\rangle_\mu$ at the time $t$. With the interaction taking place during the time interval $(0,t)$, $\psi_m(t)$ is the transition amplitude from the initial state $|m\rangle_\mu \otimes |\alpha\rangle_A$ to the state at the time $t$, $|m\rangle_t = |m\rangle_\mu \otimes |\beta(\alpha,m,t)\rangle_A$.

The involvement of $A$ in the theory, is a necessary difference between our theory and the Gisin-Percival theory; our understanding is that their state $|\psi\rangle$ can be viewed as also implicitly describing entanglement with $A$.

Thus for $A$ initially in the state $|\alpha\rangle_A$, the states $|m\rangle_t$ form the relevant basis for the time $t$. Time development during the infinitesimal time interval $(t, t+dt)$ is described by a mapping from states in the basis $|m\rangle_t$ to states in the basis $|m\rangle_{t+dt}$.

The initial state of $A$, $|\alpha\rangle_A$, is not known and it is assumed to influence the $\mu A$-interaction through the stochastic variables $\xi_m(t)$.

We shall simplify our notation and write the mappings between different $|m\rangle_t$, as if they were simply vectors in the state space for $\mu$. We let $M(t)$ be an operator in the state space of $\mu$ and we write $|m\rangle$ instead of $|m\rangle_t$, with the relations (3.7) implicitly understood.

Our basic equation then is

$$|\psi(t)\rangle = M(t)|\psi(0)\rangle = \sum_m \psi_m(t)|m\rangle. \qquad (3.9)$$

and correspondingly for the density matrix,

$$\rho(t) = M(t)\rho(0)M(t)^\dagger = \sum_{m,n} \psi_m(t)\psi_n(t)^* |m\rangle\langle n|. \qquad (3.10)$$

(Note that, according to (3.7), with respect to $A$, in (3.10) the basis for the density matrix $|m\rangle\langle n|$ is specific for each time $t$.

4. Scattering with extended process time

To obtain an equation of motion for $\mu$ (entangled with $A$), we must make an ansatz for $M(t)$. We use stochastic variables $\xi_m(\alpha, t)$, with the same means (over initial $A$ states $\alpha$) as in the introductory description in Section 2. Thus at any time $t$, for the ensemble of initial $A$ states, the stochastic variables satisfy

$$\langle\langle \xi_m \rangle\rangle = 0, \quad \langle\langle \xi_m \xi_n \rangle\rangle = 0, \quad \langle\langle \xi_m \xi_n^* \rangle\rangle = 2\delta_{mn}dt. \qquad (4.1)$$

With independent operators $L_m$, our ansatz is

$$dM(t) = \left(\sum_m L_m d\xi_m(t) - K(t)dt\right)M(t), \qquad (4.2)$$

with $K(t)^\dagger = K(t)$.

We shall determine $K(t)$, so that, in the mean, norm is preserved. Thus the mean of

$$\langle\phi|\left(\sum_m\left(L_m d\xi_m(t) + L_m^\dagger d\xi_m(t)^*\right) + \sum_{m,n}\left(L_m^\dagger d\xi_m{}^*(t)L_n d\xi_n(t)\right) - 2K(t)dt\right)|\phi\rangle = \qquad (4.3)$$
$$= \langle\phi|\left(\sum_m\left(L_m d\xi_m(t) + L_m^\dagger d\xi_m(t)^* + 2L_m^\dagger L_m dt\right) - 2K(t)dt\right)|\phi\rangle$$

should vanish for any state $|\phi\rangle$. We are using the convention that second order expressions of stochastic variables are immediately replaced by their means. The mean of (4.3) is

$$2\langle\phi|\left(\sum_m L_m^\dagger L_m - K(t)\right)|\phi\rangle dt. \qquad (4.4)$$

For this to vanish for an arbitrary state, we must have

$$K(t) = \sum_m L_m^\dagger L_m. \qquad (4.5)$$

We can then rewrite (4.2) as

$$dM(t) = \sum_m\left(L_m d\xi_m(t) - L_m^\dagger L_m dt\right)M(t). \qquad (4.6)$$

The non-normalized density matrix

$$R(t) = M(t)\rho^{(i)}M(t)^\dagger \qquad (4.7)$$

then has the differential

$$dR = dM\rho + M\rho dM^\dagger + dM\rho dM^\dagger =$$
$$= \sum_m\left(L_m R d\xi_m + RL_m^\dagger d\xi_m{}^* + \left(2L_m R L_m^\dagger - L_m^\dagger L_m R - RL_m^\dagger L_m\right)dt\right) \qquad (4.8)$$

The trace of this is the differential of the transition rate,

$$dw = w\sum_m\left(\text{Tr}(L_m\rho)d\xi_m + \text{Tr}(\rho L_m^\dagger)d\xi_m{}^*\right). \qquad (4.9)$$

Thus in the mean, the transition rate is not changed by the interaction,

$$\langle\langle dw \rangle\rangle = 0. \tag{4.10}$$

The differential for the normalized density matrix is

$$D\rho = \frac{R+dR}{w+dw} - \frac{R}{w} = \frac{1}{1+\frac{dw}{w}}\left(\frac{1}{w}dR - \frac{dw}{w}\rho\right). \tag{4.11}$$

Here we have used $D$ instead of $d$ for the differential. The reason is that $\rho + D\rho$ describes the state at the time $t+dt$. Then because of the extra factor $1+\frac{dw}{w}$ in the transition rate, the same factor appears in the distribution of states over $d\xi_m$ at the time $t+dt$ as compared to the time $t$. Therefore we have to use this factor to adjust the weight due to the number of states. This weight cancels against the denominator in the expression for $D\rho$. We have thus determined the properly weighted differential of the density matrix as

$$d\rho = \left(1+\frac{dw}{w}\right)D\rho = \frac{1}{w}dR - \frac{dw}{w}\rho. \tag{4.12}$$

It is easily calculated,

$$d\rho = \sum_m \left((L_m - \text{Tr}(L_m\rho))\rho d\xi_m + \rho(L_m^\dagger - \text{Tr}(\rho L_m^\dagger))d\xi_m^*\right) + \\ + \sum_m \left(2L_m\rho L_m^\dagger - L_m^\dagger L_m \rho - \rho L_m^\dagger L_m\right)dt. \tag{4.13}$$

Extending the stochastic $\mu A$-interaction over the time interval $(t,t+dt)$, has given us the change (4.13) in the state of $\mu$ (entangled with $A$), i.e., our equation of motion. We see that it agrees with the quantum diffusion equation (2.5).

Thus through our scattering model (4.6) within fundamental quantum mechanics, we have derived the general density-matrix equation for quantum diffusion, Equation (2.5) which is equivalent to Equation (1.2) of Gisin and Percival [1].

## 5. Measurement interaction

We consider the measurement of an observable $L = L^\dagger$ with non-degenerate eigenvalues

$$L = \sum_m l_m |m\rangle\langle m|; \quad \langle m|n\rangle = \delta_{mn}, \quad l_m{}^* = l_m, \quad l_m \neq l_n \text{ for } m \neq n. \tag{5.1}$$

Gisin and Percival [1] have described quantum diffusion (4.13) with only one operator $L_m$, namely the observable $L$ itself and only one stochastic variable $d\xi$ for each time step. We shall see that quantum diffusion leads to one of the eigenstates.

The equation of motion (4.13) for a state $\rho$ becomes

$$d\rho = (L - \mathrm{Tr}(L\rho))\rho\, d\xi + \rho(L - \mathrm{Tr}(\rho L))d\xi^* + \left(2L\rho L - L^2\rho - \rho L^2\right)dt \tag{5.2}$$

The mean of this is

$$\langle\langle d\rho_{mn}\rangle\rangle = -\rho_{mn}(l_m - l_n)^2\, dt. \tag{5.3}$$

We see that there is a general diffusion in which the diagonal elements keep their mean values. There is also a drift to zero of all non-diagonal elements. The result of this is a transition of $\mu$ into one of the eigenstates of $L$ as a result of its interaction with $A$. The state $|m\rangle\langle m|$ is approached with the relative frequency $\rho_{mm}^{(i)}$ (Born's rule).

To see the change in more detail, we can look at the diagonal elements $\rho_{mm} = p_m$ of the equation of motion (5.2),

$$dp_m = \left(l_m - \sum_n l_n p_n\right)p_m\, d\chi; \quad d\chi = d\xi + d\xi^*; \quad \langle\langle d\chi\rangle\rangle = 0, \quad d\chi^2 = 4dt. \tag{5.4}$$

Clearly

$$\langle\langle dp_m\rangle\rangle = 0, \tag{5.5}$$

and $dp_m = 0$, for $p_m = 1$ or $0$, endpoints of diffusion.

If we did have degeneracy of the eigenvalues, say if $l_m = l_r$ for $r \neq m$, then we would have

$$\frac{dp_m}{p_m} = \frac{dp_r}{p_r}, \tag{5.6}$$

i.e., the proportions between $p_m$ and $p_r$ would remain constant, and we would not get to any choice of one of them over the other.

For $\mu$ to give an imprint on $A$ of a measurement value, a more natural choice than (5.1) is to take $L_m$ as projectors on the eigenstates of the observable,

$$L_m = |m\rangle\langle m|. \tag{5.7}$$

Inserted in (4.13), this gives

$$d\rho_{mn} = \left(d\xi_m + d\xi_n^* - \sum_r \rho_{rr}(d\xi_r + d\xi_r^*) - 2(1-\delta_{mn})dt\right)\rho_{mn}. \tag{5.8}$$

The non-diagonal elemants will drift exponentially to zero. The diagonal elements undergo diffusion according to

$$d\rho_{mm} = (d\chi_m - \sum_r \rho_{rr} d\chi_r)\rho_{mm}, \tag{5.9}$$

where

$$d\chi_m = d\xi_m + d\xi_m^*; \quad \langle\langle d\chi_m \rangle\rangle = 0, \quad \langle\langle d\chi_m d\chi_n \rangle\rangle = 4\delta_{mn}dt. \tag{5.10}$$

The system $\mu$ is taken into <u>one</u> of the eigenstates of the observable, i.e., into a product state, free from any entanglement with $A$, and $A$ can be moved into a registering state related to the eigenstate of $\mu$. The state $|m\rangle\langle m|$ is again approached with the relative frequency $\rho_{mm}^{(i)}$.

In the first example with diffusion generated by $A$, the registration developmemt for $A$ can also take place but is slightly more complicated.

As with the general quantum diffusion of the previous section, the measurement processes described here, are basically ruled by linear quantum dynamics.

7. Conclusions

We have shown that the basic equations of quantum diffusion can be derived from fundamental quantum mechanics, more precisely from scattering theory of Relativistic Quantum

Mechanics, i.e., Quantum Field Theory. This means that phenomena that can be understood as quantum diffusion processes can also be understood as processes in linear quantum dynamics. No generalization is needed, nor any specially constructed phenomenological theory.

In some quantum diffusion papers, such as the referred paper by Gisin and Percival, the theory was presented as an adaptation of quantum mechanics to describe an open system, acted on by another system. We have described the theoretical connection to the basic theory in an alternative way and it agrees with their interpretation. In the comments on Gisin's and Percival's work by Bell [2] and by Whitaker [3], the closeness to linear quantum mechanics is not recognized.

Above, we have also shown how the connection, such as entanglement, between the system of interest ($\mu$) and the environment system ($A$) can be described (see, in particular, Eq. (3.7)).

For us, the decisive step in analyzing quantum effects is to use relativistic quantum mechanics rather than the non-relativistic quantum mechanics of the 1930s, which could not handle reversibility in a correct way [7].

I thank Professor K. Lindgren for his kind reading and criticism of my manuscript.

## References

1. N. Gisin and I.C. Percival, J. Phys. A Math. Gen. $\underline{25}$, 5677 (1992).

2. J.S. Bell, _Speakable and Unspeakable in Quantum Mechanics_ (2nd ed., Cambridge, U.K. 2004), 194: "The nonlinear Schrödinger equation seems to me to be the best hope for a precisely formulated theory which is very close to the pragmatic version."

3. A. Whitaker, _Einstein, Bohr and the Quantum Dilemma_ (2nd ed., Cambridge, U.K. 2006), 319-322. See also this text for earlier references.

4. See for instance L. Diósi, Journ. Phys. A $\underline{21}$, 2885 (1988).

5. K.-E. Eriksson, M. Cederwall, K. Lindgren and E. Sjöqvist, arXiv:1708.01552v2 (2017).

6. K.-E. Eriksson and K. Lindgren, Entropy <u>21</u>, 834 (2019).

7. K.-E. Eriksson, arXiv:2106.07538 (2021).